\documentclass[letterpaper,10pt,twoside]{article}
\usepackage{amsmath,amssymb,amsthm}

\newcommand{\im}{\mathrm{i}}   
\newcommand{\defeq}{:=}
\newcommand{\C}{\mathbb{C}}
\newcommand{\tens}{\otimes}    
\newcommand{\xd}{\mathrm{d}}   
\newcommand{\xD}{\mathcal{D}}  
\newcommand{\phicl}{\phi_\text{cl}}
\newcommand{\varphif}{\check{\varphi}}
\newcommand{\pop}{\kappa}
\newcommand{\hatR}{\hat{R}}
\DeclareMathOperator{\lap}{\bigtriangleup}
\newcommand{\cH}{\mathcal{H}}
\newcommand{\cS}{\mathcal{S}}
\newcommand{\cA}{\mathcal{A}}
\newcommand{\sig}{\Sigma}
\newcommand{\osig}{{\bar{\Sigma}}}

\begin{document}

\begin{titlepage}
\title{\textbf{General boundary quantum field theory: Timelike
    hypersurfaces in Klein-Gordon theory}}
\author{Robert Oeckl\footnote{email: robert@matmor.unam.mx}\\ \\
Instituto de Matem\'aticas, UNAM Campus Morelia,\\
C.P. 58190, Morelia, Michoac\'an, Mexico}
\date{15 September 2005\\9 March 2006 (v2)}

\maketitle

\vspace{\stretch{1}}

\begin{abstract}
We show that the real massive Klein-Gordon theory admits a description
in terms of states on various \emph{timelike} hypersurfaces and
amplitudes associated to regions bounded by them. This realizes
crucial elements of the general boundary framework for quantum field
theory. The hypersurfaces considered are hyperplanes on the one hand
and timelike hypercylinders on the other hand. The latter lead to
the first explicit examples of amplitudes associated with finite
regions of space, and admit no standard description in
terms of ``initial'' and ``final'' states. We
demonstrate a generalized probability interpretation in this example,
going beyond the applicability of standard quantum mechanics.

\end{abstract}

\vspace{\stretch{1}}
\end{titlepage}

\tableofcontents
\newpage
\section{Introduction}

State spaces in quantum field theory are normally associated with
spacelike hypersurfaces. What is more, in flat spacetime one
usually only considers one state space, identifying all of them
through time-translation symmetry. The restriction to spacelike
hypersurfaces has various reasons, among them the necessity to
conserve probabilities in the standard formulation of quantum
mechanics.

In contrast, we contend that this restriction is artificial. Indeed,
we have shown in \cite{Oe:timelike} that considering states on certain
timelike hypersurfaces seems to make perfect sense. The example
considered was the real massive Klein-Gordon theory and the
hypersurfaces discussed were hyperplanes. In particular, we found a
consistent vacuum state on arbitrary hyperplanes and elucidated the
nature of particles states on timelike hyperplanes.

An underlying framework into which these results may be fitted is the
so called \emph{general boundary formulation} of quantum mechanics,
also called \emph{general boundary quantum field theory}. The
foundations of this framework are laid out in the companion paper
\cite{Oe:GBQFT}. However, the present paper should be readable
independently.

The basic idea of this approach, going back to
\cite{Oe:catandclock,Oe:boundary} is that transition amplitudes make
sense not only between instances of time, but may be associated to
regions of spacetime which are not necessarily defined by a
time interval. Furthermore, the relevant state spaces are associated
to the boundary hypersurfaces of these regions.

In the present article we continue the development of Klein-Gordon
theory in this framework. The key new example of hypersurface we
consider is an infinite timelike hypercylinder. More precisely,
consider a sphere in space and extend it over all of time. This is
what we call the \emph{hypercylinder}. The spacetime regions of
interest in this context are the solid hypercylinder as well as a
solid hypercylinder with another one cut out.

Using the Schr\"odinger representation combined with the Feynman path
integral (explained in Section~\ref{sec:basic}) we work out state
spaces of wave functions, field propagators, vacuum states and
particle states. We show that all these structures are consistent in
the sense of general boundary quantum field theory. In doing so, we
start by reviewing spacelike hyperplanes (Section~\ref{sec:sl}) to
clarify the approach, then move to recall and elaborate on the general
hyperplane case (Section~\ref{sec:gh}). The main technical results are
obtained in the treatment of the hypercylinder case
(Section~\ref{sec:hcl}), with the novel situation of an amplitude
associated to a region of spacetime with \emph{connected}
boundary. This case requires a genuinely new interpretation and the
general boundary formulation shows its full force here.

After the technical part we move to a discussion of the interpretation
(Section~\ref{sec:interpret}). This consists firstly of a discussion
of the meaning of particle states on timelike hypersurfaces. In
particular, we elaborate on the fact that particles within a state
acquire the property of being incoming or outgoing \emph{individually}.
Secondly, we apply the probability interpretation proposed in
\cite{Oe:GBQFT} to the case of the solid hypercylinder. Due to the
connectedness of the boundary an amplitude is to be evaluated on a
\emph{single} state. While this is out of the range of
applicability of standard quantum mechanics, the proposed
interpretation yields a physically fully satisfactory answer.
We close with a few remarks (Section~\ref{sec:concl}).

\section{Generalized Schr\"odinger-Feynman approach}
\label{sec:basic}

Our aim is to show how the Klein-Gordon theory admits a generalized
description in terms of a general boundary quantum field theory. More
precisely, we wish to show how it admits states on hypersurfaces that
are not necessarily spacelike and amplitudes associated to spacetime
regions which are not necessarily time intervals.
It turns out that to this end the Schr\"odinger representation (where
states are wave functions) combined with the Feynman path integral
\cite{Fey:stnrqm} are
particularly suitable \cite{Oe:boundary}. Since the former is not
usually employed in quantum field theory, we refer to the review
\cite{Jac:schroedinger}.

We start by introducing the basic structures and their heuristic
definition. This includes state spaces, inner products, propagators
etc. Most of these exist already in the standard formalism, but
will be defined here in a generalized way. Furthermore, consistency
conditions will be formulated that these structures must
satisfy. These conditions may be derived directly from the axiomatic
framework presented in the companion paper \cite{Oe:GBQFT}, see also
Section~7 of that paper.
Again, these generalize consistency conditions of the standard quantum
mechanical framework, including unitarity, composability of time
evolutions etc.
That this generalization really
makes sense is the subject of the remaining part of this article.

\subsection{States and amplitudes}

The basic spacetime objects we will need are hypersurfaces $\sig$ and
regions (4-submanifolds) $M$ in Minkowski space. The former generalize
spacelike hypersurfaces and the latter generalize regions of
time-evolution between them.

On a given hypersurface $\sig$ we consider the space of 
field configurations, which we denote by $K_\sig$. Since we are
dealing with a theory of one real scalar field this is basically the
space of real-valued functions on $\sig$.\footnote{We will later see
that this naive definition is not always correct. However, it will
suffice here for the intuitive picture.}
The
Schr\"odinger representation prescribes now
that states are
complex-valued wave functions on this configuration space. Thus, we
associate with $\sig$ its space of states (wave functions) which we
denote $\cH_\sig$. This state space carries 
an inner product, defined through
an integral over field configurations,
\begin{equation}
  \langle\psi,\psi'\rangle_\sig\defeq\int_{K_\sig}\xD\varphi\,
  \overline{\psi(\varphi)}\psi'(\varphi) .
\label{eq:schrip}
\end{equation}

Note the following important property of state spaces. Take a
hypersurface $\sig$ which is the disjoint union of two hypersurfaces
$\sig_1\cup\sig_2$. The field configurations on $\sig$ are then
obviously pairs of field configurations on $\sig_1$ and on
$\sig_2$, i.e., $K_\sig=K_{\sig_1}\times K_{\sig_2}$. Thus, wave
functions on $\sig$ can be expanded into products of wave functions on
$\sig_1$ and $\sig_2$, i.e.,
$\cH_\sig=\cH_{\sig_1}\tens\cH_{\sig_2}$.

We will make use of another structure on hypersurfaces:
orientation. From now on we will
think of each hypersurfaces as oriented, i.e., each hypersurface has
a chosen ``side''. For a given oriented hypersurface $\sig$, we denote
its oppositely oriented version by $\osig$, i.e., using an
over-bar. Although the associated state spaces $\cH_\sig$ and
$\cH_\osig$ are canonically identical (indeed, so far we have not
distinguished between them), their physical meaning must be
distinguished in the following. However, there is a map that tells us
which state on $\osig$ corresponds physically to a given state on
$\sig$. This map, denoted by $\iota_\sig:\cH_\sig\to\cH_\osig$ is
given by the complex conjugation of wave functions,
\begin{equation}
 (\iota_\sig(\psi))(\varphi)\defeq\overline{\psi(\varphi)}\quad\forall
 \psi\in\cH_\sig,\varphi\in K_\sig .
\label{eq:conj}
\end{equation}

Let $M$ be a region with boundary $\sig$. The hypersurface $\sig$ is
oriented by choosing the ``outside'' of $M$. We define the
\emph{amplitude} associated with $M$ as the map
$\rho_M:\cH_\sig\to\C$, associating with each wave function a complex
number as follows,
\begin{gather}
 \rho_M(\psi)\defeq\int_{K_\sig}\xD\varphi\, \psi(\varphi) Z_M(\varphi)
 \quad\forall \psi\in\cH_\sig,
\label{eq:ampl}\\
 Z_M(\varphi)\defeq\int_{K_M, \phi|_\sig=\varphi}\xD\phi\, e^{\im
 S_M(\phi)}\quad\forall \varphi\in K_\sig .
\label{eq:prop}
\end{gather}
The second integral is the Feynman path integral over ``all field
configurations'' $\phi\in K_M$ in the
region $M$ that reduce to $\varphi$ on the boundary $\sig$. $S_M$ is the
action integral over the region $M$. The quantity $Z_M(\varphi)$ is
called the \emph{field propagator}. It formally looks like a wave
function, but might not be normalizable with respect to the inner
product (\ref{eq:schrip}).

Consider a region $M$ with boundary $\sig$ consisting of the disjoint
union of two components $\sig_1$ and $\sig_2$. Then, the amplitude
$\rho_M:\cH_{\sig_1}\tens\cH_{\sig_2}\to\C$ induces a map
$\tilde{\rho}_M:\cH_{\sig_1}\to\cH_{\osig_2}$ via
\begin{equation}
 (\tilde{\rho}_M(\psi))(\varphi')
 =\int_{K_{\sig_1}}\xD\varphi\, \psi(\varphi) Z_M(\varphi,\varphi')
 \quad\forall \psi\in\cH_{\sig_1},\varphi'\in K_{\osig_2} .
\label{eq:indevol}
\end{equation}
Note that the orientation reversal on $\sig_2$ comes from the fact
that the Hilbert space $\cH_{\sig_2}$ must be dualized when moved from
the domain of $\rho_M$ to the image. But 
the inner product (\ref{eq:schrip}) together with the conjugation
(\ref{eq:conj}) make $\cH_{\osig_2}$ precisely into the dual space of
$\cH_{\sig_2}$. The physical meaning of $\tilde{\rho}_M$ is that (in
suitable circumstances) we may think of it as describing the evolution
(not necessarily in time) of the system from $\sig_1$ to $\osig_2$.

In particular, suppose we have two spacelike hypersurfaces $\sig_1$
and $\osig_2$ and call the intermediate region $M$. Then,
$\tilde{\rho}_M(\psi)$ is the time-evolved wave function on $\osig_2$
for the initial wave function $\psi$ on $\sig_1$. Note here that
$\sig_1$ and $\osig_2$ have the same orientation with respect to the
time direction, namely with their ``past'' side selected. Indeed, the
reason why the orientation of hypersurfaces does not appear explicitly
in the standard formulation is that all spacelike hypersurfaces can be
coherently oriented in the same way, using the time direction.

We now formulate what a \emph{unitarity} evolution
(i.e., an evolution preserving the inner product) means. One can check
that for $\tilde{\rho}_M$ to be unitary the propagator must satisfy the
following formal condition,
\begin{equation}
 \int_{K_{\sig_2}}\xD\varphi_2\, \overline{Z_M(\varphi_1,\varphi_2)}
 Z_M(\varphi_1',\varphi_2)=\delta(\varphi_1,\varphi_1')\quad
 \forall\varphi_1,\varphi_1'\in K_{\sig_1}.
\label{eq:propunit}
\end{equation}

Finally, consider a region $M_1$ with boundary consisting of
disconnected components $\sig_1$ and $\sig$ and a region $M_2$ with
boundary consisting of disconnected components $\osig$ and $\sig_2$
such that $M_1$ and $M_2$ may be glued along $\sig$ to form a new
region $M$. We then want that the evolution associated with $M$ is the
composition of the evolutions associated with $M_1$ and $M_2$. That is,
$\tilde{\rho}_{M_1\cup
  M_2}=\tilde{\rho}_{M_2}\circ\tilde{\rho}_{M_1}$. In terms of
propagators this means,
\begin{equation}
 \int_{K_{\osig}}\xD\varphi\, Z_{M_1}(\varphi_1,\varphi)
  Z_{M_2}(\varphi,\varphi_2) = Z_{M_1\cup M_2}(\varphi_1,\varphi_2)
  \quad\forall\varphi_1\in K_{\sig_1},\varphi_2\in
  K_{\osig_2}.
\label{eq:pintcomp}
\end{equation}
If we write the propagator in terms of the path integral
(\ref{eq:prop}) the validity of (\ref{eq:pintcomp}) becomes
obvious.\footnote{However, it becomes much less obvious when going
  beyond the naive picture presented here. We will see this in the
  following.}

\subsection{Vacuum}

We postulate that there is a distinguished vacuum wave function on
each hypersurface $\sig$, denoted $\psi_{\sig,0}$. We require the
vacuum to satisfy certain properties (the vacuum axioms of
\cite{Oe:GBQFT}). This is, firstly, the compatibility with
conjugation. This means that the vacuum wave function on a
hypersurface is complex conjugate to the vacuum on the hypersurface
with the opposite orientation. Formally,
\begin{equation}
 \psi_{\osig,0}(\varphi)=(\iota_\sig(\psi_{\sig,0}))(\varphi)
 =\overline{\psi_{\sig,0}(\varphi)}\quad\forall\varphi\in K_\sig .
\label{eq:vacconj}
\end{equation}

Another property of the vacuum state we expect is that for a
hypersurface $\sig$ consisting of disjoint hypersurfaces $\sig_1$ and
$\sig_2$ the vacuum wave function should be the product of the
individual vacuum wave functions, i.e.,
\begin{equation}
 \psi_{\sig,0}(\varphi_1,\varphi_2)
 =\psi_{\sig_1,0}(\varphi_1)\psi_{\sig_2,0}(\varphi_2)\quad
 \forall \varphi_1\in K_{\sig_1}, \varphi_2\in K_{\sig_2} .
\label{eq:vacprod}
\end{equation}

We also want the vacuum states to be normalized, i.e.,
\begin{equation}
 \int_{K_\sig}\xD\varphi\, |\psi_{\sig,0}(\varphi)|^2=1 .
\label{eq:vacnorm}
\end{equation}

Finally, the amplitude of the vacuum state should be unity. Suppose
$M$ is a region with boundary $\sig$, then
\begin{equation}
 \rho_M(\psi_{\sig,0})
 =\int_{K_\sig}\xD\varphi\, \psi_{\sig,0}(\varphi) Z_M(\varphi)= 1 .
\label{eq:vacampl}
\end{equation}

As can be shown \cite{Oe:GBQFT} these properties imply that the vacuum
is preserved under evolution in the sense described above. That is,
evolving from a hypersurface $\sig_1$ to a hypersurface $\osig_2$ via
a region $M$ with evolution map 
$\tilde{\rho}_M:\cH_{\sig_1}\to\cH_{\osig_2}$, the vacuum satisfies
\begin{equation}
 \psi_{\osig_2,0}(\varphi_2)
 =(\tilde{\rho}_M(\psi_{\sig_1,0}))(\varphi_2)=
 \int_{K_{\sig_1}}\xD\varphi_1\, \psi_{\sig_1,0}(\varphi_1)
 Z_M(\varphi_1,\varphi_2)\quad\forall\varphi_2\in K_{\sig_2}.
\label{eq:vacevol}
\end{equation}
In the context of parallel equal-time hyperplanes this is the
standard notion of time-evolution invariance of the vacuum.

\section{Spacelike hyperplanes}
\label{sec:sl}

In this section we review standard elements of the Schr\"odinger
representation of the Klein-Gordon theory. This will make the later
generalization and its meaning more transparent.
We follow here a presentation close to \cite{Oe:timelike}.
Hypersurfaces
are here equal-time hyperplanes and regions are time intervals
extended over all of space. We use time translation symmetry to
identify the spaces of wave functions associated to
all (past oriented) equal-time hyperplanes. (See \cite{Oe:GBQFT} for
a more detailed discussion of this identification.)

We start by recalling elementary features of the classical
Klein-Gordon theory of a real scalar field $\phi$ with mass $m$ in
Minkowski space. The equations of motion are given by the Klein-Gordon
equation, $(\Box +m^2)\phi=0$, with $\Box\defeq
\partial_0^2-\sum_{i\ge 1}\partial_i^2$. The action
on a region $M$ of Minkowski space is given by
\[
 S_M(\phi)=\frac{1}{2}\int_M \xd^4 x\, \left((\partial_0 \phi)
 (\partial_0 \phi)- \sum_{i\ge 1} (\partial_i \phi)(\partial_i \phi)
 -m^2\phi^2\right) .
\]
We may rewrite this as follows,
\begin{equation}
 S_M(\phi)=-\frac{1}{2}\int_M \xd^4 x\, \phi (\Box + m^2) \phi
  +\frac{1}{2}\int_{\partial M} \xd^3 x\,
 \phi (n_0\partial_0-\sum_{i\ge 1} n_i \partial_i) \phi .
\label{eq:actionbdy}
\end{equation}
Here $n_i$ is the local euclidean normal vector to the boundary
$\partial M$ of $M$ pointing \emph{outwards}. Note that by the
equations of motion, the action applied to a solution reduces to the
boundary term.

\subsection{Propagator}

Consider two
instants in time $t$ and $t'$. The time interval $[t,t']$ defines
a region in Minkowski space in the obvious way. Its boundary has two
connected components associated with the two
instants of time. 
Using the variational principle that determines the equations
of motions together with the fact that the action is quadratic we can
evaluate the field propagator (\ref{eq:prop}) for this
region. Choosing a classical solution $\phicl$
matching the boundary data at $t$ and $t'$ and shifting the integration
variable yields
\begin{equation}\begin{split}
 Z_{[t,t']}(\varphi,\varphi') & =
 \int_{\phi|_{t}=\varphi, \phi|_{t'}=\varphi'} \xD \phi\,
  e^{\im S_{[t,t']}(\phi)} \\
 & = \int_{\phi|_{t}=0, \phi|_{t'}=0} \xD \phi\,
  e^{\im S_{[t,t']}(\phicl+\phi)} \\
 & = N_{[t,t']} e^{\im S_{[t,t']}(\phicl)} ,
\label{eq:propcl}
\end{split}\end{equation}
where $\Delta\defeq t'-t$.
The normalization factor is formally given by
\[
N_{[t,t']} = \int_{\phi|_{t}=0, \phi|_{t'}=0} \xD \phi\,
  e^{\im S_{[t,t']}(\phi)} .
\]
Using (\ref{eq:actionbdy})
the evaluation of the action on a classical
solution reduces to a boundary integral. In the present case
this is
\begin{equation}
 S_{[t,t']}(\phicl)=\frac{1}{2}\int_{t'} \xd^3 x\, \phicl(t',x)
 \partial_0 \phicl(t',x)
 - \frac{1}{2}\int_{t} \xd^3 x\, \phicl(t,x) \partial_0 \phicl(t,x) .
\label{eq:bdyaction}
\end{equation}
We split the classical solution into positive and negative energy
components
\begin{equation}
\phicl(t,x)=e^{-\im\omega t}\varphi^+(x)+e^{\im\omega t}\varphi^-(x) ,
\label{eq:edec}
\end{equation}
where $\omega$ is the operator
\[\omega\defeq\sqrt{-\sum_{i\ge 1}
\partial_i^2+m^2} .
\]
Inserting this decomposition into (\ref{eq:bdyaction}) and inverting
the formal linear transformation
\[
 \begin{pmatrix}\varphi \\ \varphi'\end{pmatrix}
 =\begin{pmatrix}e^{-\im\omega t} & e^{\im\omega t}\\
 e^{-\im\omega t'} & e^{\im\omega t'}\end{pmatrix}
 \begin{pmatrix}\varphi^+ \\ \varphi^- \end{pmatrix}
\]
yields the field propagator (see \cite{Oe:boundary}),
\begin{equation}
 Z_{[t,t']}(\varphi,\varphi')=N_{[t,t']} \exp
 \left(-\frac{1}{2}\int\xd^3 x
 \begin{pmatrix}\varphi & \varphi' \end{pmatrix} W_{[t,t']}
 \begin{pmatrix}\varphi \\ \varphi' \end{pmatrix}\right) .
\label{eq:proptbdy}
\end{equation}
The operator-valued matrix $W_{[t,t']}$ is given by
\[
 W_{[t,t']}=\frac{-\im\omega}{\sin\omega\Delta}
 \begin{pmatrix}\cos\omega\Delta & -1 \\
  -1 & \cos\omega\Delta \end{pmatrix} .
\]

\subsection{Vacuum}

To obtain the vacuum wave function consider the Gaussian ansatz
\begin{equation}
 \psi_{0}(\varphi)=C \exp\left(-\frac{1}{2}\int \xd^3 x\, \varphi(x)
(A \varphi)(x)\right)
\label{eq:vacansatz}
\end{equation}
for an unknown operator $A$ and a normalization constant $C$. Imposing
the time-evolution invariance (\ref{eq:vacevol}) yields the equation
$A^2=\omega^2$. We choose $A=\omega$, following the standard
conventions. At the same time, this fixes the normalization factor
$N_{[t,t']}$ to be formally
\begin{equation}
 N_{[t,t']}^{-1}=\int\xD\varphi \exp\left(-\frac{1}{2}\int\xd^3 x\,
 \varphi(x)\frac{\omega \exp(\im\omega\Delta)}
 {\im\sin(\omega\Delta)}\varphi(x)\right) .
\label{eq:propnorm}
\end{equation}
One can check that with this normalization, the propagator satisfies
both the unitarity condition (\ref{eq:propunit}) and the composition
property (\ref{eq:pintcomp}).

The normalization condition (\ref{eq:vacnorm}) for the vacuum fixes the
factor $C$ (up to a phase),
\begin{equation}
 |C|^{-2}= \int \xD\varphi\, \exp\left(-\frac{1}{2}\int \xd^3 x\, 
 \varphi(x)(2\omega \varphi)(x)\right).
\label{eq:svacnorm}
\end{equation}
We choose $C$ to be real. We will come back to the reason for this
later. Note that this implies that the vacuum wave function is real
and thus (implementing (\ref{eq:vacconj})) the same for both
orientations of an equal-time hyperplane.

The property (\ref{eq:vacprod}) is now a definition that only comes
into force if we consider unions of different equal-time
hyperplanes. The unit amplitude property (\ref{eq:vacampl}) follows
from the other properties already implemented.

\subsection{Particle states}

The particle states can be found by use of suitable creation and
annihilation operators (see \cite{Jac:schroedinger}) or through
an expansion of the propagator in terms of eigenstates. We will not
enter into the details here.

Using the Fourier transform,
\[
 \varphif(p)=2 E \int\xd^3 x\, e^{\im p x} \varphi(x),\quad
 \varphi(x)=\int\frac{\xd^3 p}{(2\pi)^3 2E} e^{-\im p x}
 \varphif(p)
\]
the one-particle wave function of momentum $p$ is given by
\begin{equation}
 \psi_p(\varphi)=\varphif(p)\psi_0(\varphi) .
\label{eq:onepstate}
\end{equation}
These wave functions have a distributional normalization, given by
\begin{equation}
 \langle \psi_p , \psi_{p'}\rangle = (2\pi)^3 2 E \delta^3(p-p') .
\end{equation}
The two-particle state with momenta $p$ and $p'$ is given by the wave
function
\begin{equation}
 \psi_{p,p'}(\varphi)=\left(
  \varphif(p)\varphif(p')
  -(2\pi)^3 2 E \delta^3(p+p')\right)\psi_0(\varphi) .
\label{eq:twopstate}
\end{equation}
The $n$-particle wave function is a polynomial of degree $n$ in Fourier
transforms times the vacuum wave function. To obtain it explicitly for
given momenta $p_1,\dots,p_n$, one may project the wave function
\[
 \varphif(p_1)\cdots\varphif(p_n)\psi_0(\varphi)
\]
to the orthogonal complement of the spaces of states with less than
$n$ particles.

In contrast to the vacuum, particle wave functions are generically not
real. Thus, (\ref{eq:conj}) prescribes that a given particle wave
function change explicitly to its
complex conjugate when considered on an equal-time hyperplane with
opposite orientation. Since we have taken ``past-orientation'' as
standard, we denote a wave function $\psi$, when specifying the same
state but with ``future-orientation'' by $\bar{\psi}$.

The transition amplitude from an $n$-particle state
$\psi_{p_1,\dots,p_n}$ at time $t_1$ to to an $m$-particle state
$\psi_{q_1,\dots,q_m}$ is given
by the amplitude function (\ref{eq:ampl}) and
evaluates to
\begin{multline*}
 \rho_{[t,t']}(\psi_{p_1,\dots,p_n}\tens\bar{\psi}_{q_1,\dots,q_m})\\
 = \int_{K_t\times K_{t'}}\xD\varphi\xD\varphi'\,
  \psi_{p_1,\dots,p_n}(\varphi)\overline{\psi_{q_1,\dots,q_m}(\varphi')}
  Z_{[t,t']}(\varphi,\varphi')\\
 = \delta_{n,m} e^{-\im \sum_{i=1}^n \Delta E_i}
  \sum_{\sigma\in S_n}\prod_{i=1}^{n} (2\pi)^3 2E_i
  \delta^3(p_i-q_{\sigma(i)}) .
\end{multline*}
The sum runs over all permutations $\sigma$ of $n$ elements.

Using (\ref{eq:indevol}) the notation for the
amplitude may be brought into the more conventional form
\begin{equation}
 \rho_{[t,t']}(\psi_{p_1,\dots,p_n}\tens\bar{\psi}_{q_1,\dots,q_m})
 =\langle \psi_{q_1,\dots,q_m},\tilde{\rho}_{[t,t']}
 (\psi_{p_1,\dots,p_n})\rangle,
\label{eq:samplbk}
\end{equation}
which recovers the usual bra-ket notation.

\section{General hyperplanes}
\label{sec:gh}

We now generalize from equal-time hyperplanes to arbitrary
hyperplanes in Minkowski space \cite{Oe:timelike}. We start with a
particular timelike hyperplane.

Consider the hyperplane aligned with the time axis
and spanned by coordinate directions $(t,x_2,x_3)$. We will denote the
coordinates $x_2,x_3$ collectively by $\tilde{x}$.
A crucial difference to
the spacelike case arises as follows. When
considering field
configurations in the sense of Section~\ref{sec:basic} we have to keep
in mind that a field configuration together with the conjugate
momentum should correspond to a classical solution. If we consider
configuration and momentum on a spacelike
hyperplane, then clearly, a configuration can be essentially any real
function on the hyperplane. However, on a timelike hyperplane not
every real function extends to a classical solution. It is easy to
see that the space of configurations that do extend is the
subspace of the space of ``all'' configurations on which the square of
the operator
\[
 \kappa_1\defeq\sqrt{-\partial_0^2+\sum_{i\ge 2} \partial_i^2-m^2}
\]
has non-negative eigenvalues. (Thus, $\kappa_1$ itself is well defined
on this space.) We call this the \emph{physical}
configuration space.

\subsection{Timelike propagator}

The first object
of interest in this context is the propagator, say for the spacetime
region defined by the interval $[x_1,x_1']$. It was shown in
\cite{Oe:timelike} that it can be calculated along lines entirely
analogous to the equal-time hyperplane case. Namely, the path integral
can be evaluated with the help of a classical solution matching the
boundary data. This involves using the boundary form
(\ref{eq:actionbdy}) of the action evaluated on the classical
solution. This time, the step analogous to the decomposition
(\ref{eq:edec}) is the
decomposition in terms of solutions with positive versus negative
momentum in the $x_1$-direction,
\[
 \phi(t,x)=e^{\im \pop_1 x_1} \varphi^+(t,\tilde{x})
  +e^{-\im \pop_1 x_1} \varphi^-(t,\tilde{x}) .
\]
This leads to the result,
\begin{equation}
 Z_{[x_1,x_1']}(\varphi,\varphi')=N_{[x_1,x_1']}
 \exp\left(-\frac{1}{2}\int  \xd t\,\xd^2\tilde{x}
 \begin{pmatrix}\varphi & \varphi' \end{pmatrix} W_{[x_1,x_1']}
 \begin{pmatrix}\varphi \\ \varphi' \end{pmatrix}\right) ,
\label{eq:propsbdy}
\end{equation}
with $\Delta\defeq |x_1'-x_1|$.
The operator-valued matrix $W_{[x_1,x_1']}$ is given by
\[
 W_{[x_1,x_1']}=\frac{\im\pop_1}{\sin\pop_1\Delta}
 \begin{pmatrix}\cos\pop_1\Delta & -1 \\
  -1 & \cos\pop_1\Delta \end{pmatrix} .
\]

\subsection{Timelike vacuum}

To determine the vacuum state we make the ansatz analogous to
(\ref{eq:vacansatz}), i.e., using a Gaussian bilinear form with
undetermined operator $A$. Explicit computation yields
$A^2=\pop_1^2$. We choose $A=\pop_1$, justifying this choice
later. Note that this also fixes the normalization factor
$N_{[x_1,x_1']}$ appearing in the propagator to
\[
 N_{[x_1,x_1']}^{-1}=\int \xD \varphi \exp\left(-\frac{1}{2}\int
 \xd t\,\xd^2\tilde{x}\, \varphi
 \frac{\im\kappa_1\exp(-\im\kappa_1\Delta)}{\sin \kappa_1\Delta}
 \varphi \right) .
\]
One can check now that the unitarity condition (\ref{eq:propunit}) as
well as the composition property (\ref{eq:pintcomp}) are
satisfied.

Especially the latter fact is actually surprising and
merits a remark. The composition property (\ref{eq:pintcomp}), when
introduced in Section~\ref{sec:basic} seemed obviously correct.
This was for the
simple reason that the propagator (\ref{eq:prop}) is a path integral,
which by its very meaning should be sliceable into pieces if one
integrates over all intermediate configurations. However, since the
path integral is over ``all'' configurations (real functions) in
spacetime, the configurations to be integrated over on the boundary
between slices should be ``all'' configurations (real functions). This
is not what we are doing. We are only integrating over the physical
configurations as explained above. That the composition rule holds
nevertheless is thus a non-trivial fact
(in contrast to the spacelike case).

The normalization condition (\ref{eq:vacnorm}) on the vacuum yields
now
\[
 |C|^{-2}=\int \xD \varphi \exp\left(-\frac{1}{2}\int
 \xd t\,\xd^2\tilde{x}\, \varphi 2 \kappa_1
 \varphi \right) .
\]
One may argue that this yields the same $|C|$ as (\ref{eq:svacnorm})
by putting the two configurations spaces and their measures into
correspondence.
There is a novel aspect concerning the phase of $C$ here. Recall from
(\ref{eq:vacconj}) that
the vacuum wave function on an oppositely oriented hyperplane must be
given by the complex conjugate. However, we may use a spatial rotation
to transform a timelike hyperplane into itself, but with opposite
orientation. Thus, rotating the vacuum this way and requiring equality
with the complex conjugate yields the \emph{condition} that the vacuum
wave function must be real. This implies $C$ to be real.\footnote{One
might try to use a similar argument in the spacelike case using a
time reflection transformation. However, this transformation is not
connected to the identity of the Poincar\'e group and might thus not
necessarily be expected to leave the vacuum invariant.}

\subsection{Timelike particle states}

After having convinced ourselves that the structures defined have all
properties listed in Section~\ref{sec:basic} we move to consider
particle states.
Since a state is a wave function on physical configurations on the
hyperplane, a basis of one-particle states may be characterized by the
Fourier modes in this hyperplane. In the standard (spacelike) case
these are labeled by $3$-momentum. In the present (timelike) case these
are labeled by the (possibly negative) energy and the momentum in the
$\tilde{x}$-directions. We will consider here merely the formal
properties of particle states, postponing a discussion of their
meaning to Section~\ref{sec:interpret}.

Since we set the energy variable to be positive, $E>0$, we distinguish
the actual sign of the energy by an index $\pm$, using a
Fourier transform of the form
\begin{equation}
\varphif^{\pm}(E,\tilde{p}) \defeq 2p_1 \int \xd t\,\xd^2 \tilde{x}\,
 e^{\pm\im  (E t - \tilde{p} \tilde{x})} \varphi(t,\tilde{x}) .
\label{eq:tfourier}
\end{equation}
The one-particle state of energy $E$ (or $-E$) and 2-momentum
$\tilde{p}$ is given by the wave function
\begin{equation}
\psi_{E,\tilde{p}}^{\pm}(\varphi)=\varphif^{\pm}(E,\tilde{p})
 \psi_0(\varphi) .
\label{eq:tstate}
\end{equation}
Its eigenvalue under ``spatial evolution'' from $x_1$ to $x_1'$ is
given by $\exp(\im \Delta p_1)$, where $p_1$ is the positive square
root $p_1=\sqrt{E^2-\tilde{p}^2-m^2}$. The inner product of
one-particle states is given by the distribution
\begin{equation}
 \langle \psi_{E,\tilde{p}}^a, \psi_{E',\tilde{p}'}^{a'}\rangle
 = (2\pi)^3 2 p_1 \delta_{a,a'}\delta(E-E')
 \delta^2(\tilde{p}-\tilde{p}') .
\end{equation}
The complex conjugation of the wave function associated with a change
of orientation of the hyperplane simply changes the sign index,
\begin{equation}
 \overline{\varphif^{\pm}(E,\tilde{p})
 \psi_0(\varphi)}=\varphif^{\mp}(E,\tilde{p})
 \psi_0(\varphi) .
\label{eq:conjtl}
\end{equation}

Multi-particle states are formed in analogy to the spacelike case,
namely by starting with a monomial in (\ref{eq:tfourier}) times the
vacuum wave function and then projecting out the components in
subspaces of lower particle number. For example, the two-particle
state takes the form
\begin{multline}
 \psi_{(E,\tilde{p}),(E',\tilde{p}')}^{a,a'}(\varphi)\\
 =\left(\varphif^{a}(E,\tilde{p})
 \varphif^{a'}(E',\tilde{p}')
  -(2\pi)^3 2 p_1 \delta_{a,-a'}\delta(E-E')
 \delta^2(\tilde{p}-\tilde{p}')\right)\psi_0(\varphi) .
\label{eq:twoptstate}
\end{multline}

Similarly to the spacelike case, we may identify state spaces
associated to parallel hyperplanes by using a spatial translation
symmetry. We may thus write amplitudes between such parallel
hyperplanes. For example, the one-particle to one-particle amplitude
is
\begin{multline}
 \rho_{[x_1,x_1']}(\psi_{E,\tilde{p}}^a
 \tens\psi_{E',\tilde{p}'}^{a'})\\
 = \int_{K_{x_1}\times K_{x_1'}}\xD\varphi\xD\varphi'\,
\psi_{E,\tilde{p}}^a(\varphi)\psi_{E',\tilde{p}'}^{a'}(\varphi')
  Z_{[x_1,x_1']}(\varphi,\varphi')\\
 = e^{\im \Delta p_1} (2\pi)^3 2 p_1 \delta_{a,-a'}\delta(E-E')
 \delta^2(\tilde{p}-\tilde{p}') .
\label{eq:tampl}
\end{multline}
Note that no explicit complex conjugation appears here, since we have
chosen the wave functions with respect to the orientations of the
carrying hyperplanes as boundaries of the enclosed region.

As in the spacelike case we may rewrite the amplitude in a form
analogous to (\ref{eq:samplbk}), reminiscent of the bra-ket notation.
However, this is no longer very useful. In contrast to the spacelike
case, rotational symmetry prevents a
consistent orientation of all timelike
hypersurfaces from the outset. Thus, none of the two possible ways of
of writing the amplitude (\ref{eq:tampl}) in the form
(\ref{eq:samplbk}), arising from the two possible orientations is
preferred.

\subsection{General vacuum}

We now turn to arbitrary hyperplanes in Minkowski space. Such a
hyperplane is either spacelike, timelike or null. In the first case we
can obtain it by a Poincar\'e transformation from the equal-time
hyperplane considered in Section~\ref{sec:sl}. In the second case we
can obtain it by a Poincar\'e transformation from the timelike
hyperplane considered above. Since the theory is fully Poincar\'e
covariant, state spaces, propagators etc.\ can all be obtained in a
straightforward way by the induced transformations.
We will not detail the results here, but limit ourselves to
one object which indicates the consistency of the present
approach in a surprising way. This is the vacuum wave function.

Consider an arbitrary hyperplane.
Suppose the angle between time axis and the euclidean normal vector to
the hyperplane is given by $\alpha$.
Since the effect of translations and of spatial rotations is
straightforward and uninteresting, we
translate and rotate our coordinate system such that the hyperplane in
question is spanned by coordinate directions $(s,x_2,x_3)$. Here $s$
is a Euclidean coordinate along
the hyperplane such that $x_1=s\cos\alpha$ and $t=s\sin\alpha$.

It was shown in \cite{Oe:timelike} using suitable Lorentz boosts,
that both, in the spacelike as well as in the timelike case
the vacuum wave function on the hyperplane may be written as
\begin{equation}
 \psi_0(\varphi)=C \exp\left(-\frac{1}{2}\int \xd s\, \xd^2
 \tilde{x}\, \varphi(s,\tilde{x}) (\tau \varphi)(s,\tilde{x})\right) ,
\label{eq:vacgeneral}
\end{equation}
where $\tau$ is an operator.
What is more, the operator $\tau$ takes a simple form which covers
\emph{both}, the spacelike \emph{and} the timelike case,
\begin{equation}
 \tau = \sqrt{-\partial_s^2+\cos 2\alpha
  \left(-\sum_{i\ge 2}\partial_i^2+m^2\right)} .
\end{equation}
It is easy to see that for $\alpha=0$ we recover $\omega$ and for
$\alpha=\pi/2$ we recover $\pop_1$. (This fixes the choice of sign
encountered earlier.) Remarkably, however, the vacuum
wave function depends smoothly on the angle $\alpha$, not only in the
intervals $0\le\alpha<\pi/4$ (spacelike) and $\pi/4<\alpha\le\pi/2$
(timelike), but even at and near $\pi/4$ (null). On the one hand this
indicates that our separate treatments of the spacelike and timelike
cases are indeed consistent with each other. On the other hand this
suggests that even states on null hyperplanes may make sense.

\section{The hypercylinder}
\label{sec:hcl}

In this section
we consider hypersurfaces which are infinite hypercylinders
(and hyperspheres) in the following sense. Consider a sphere of radius
$R$ in space. Take the hypersurface formed by the extension of this
sphere over all of time in Minkowski space. We will call this simply
the \emph{hypercylinder} of radius $R$.
At the same time we
consider regions of spacetime given by a solid hypercylinder $B_R$.
Furthermore, we consider the solid hypercylinder (of radius $\hatR$)
with a smaller solid hypercylinder (of radius $R$) cut out,
denoting this region by $B_{[R,\hatR]}$.

\subsection{Coordinates, classical solutions etc.}

We use spherical coordinates in space, parametrized by angles
$\theta\in[0,\pi[$ and $\phi\in[0,2\pi[$ and the radius 
$r\in[0,\infty[$. Concretely, we use the coordinate transformations
$x_1 =r \sin\theta \cos\phi$, $x_2 =r \sin\theta \sin\phi$,
$x_3 =r \cos\theta$.
The Laplace operator $\lap\defeq\sum_{i\ge 1} \partial_i^2$ takes the
form $\lap=\lap_r+\lap_\Omega$ with
\[
 \lap_r\defeq \frac{2}{r}\partial_r + \partial_r^2\quad
 \text{and}\quad
 \lap_\Omega \defeq \frac{\cos\theta}{r^2 \sin\theta}\partial_\theta
  +\frac{1}{r^2 \sin\theta}\partial_\theta^2
  +\frac{1}{r^2 \sin\theta}\partial_\phi^2 .
\]

We can expand solutions of the equations of motions in terms of
spherical harmonics via
\begin{equation}
 \phi(t,r,\Omega)=\int \xd E \sum_{l=0}^\infty\sum_{m=-l}^l
 \alpha_{l,m}(E) e^{-\im E t} f_l(p r) Y^m_l(\Omega) .
\label{eq:sphsol}
\end{equation}
The integral over $E$ is constrained to $|E|\ge m$ and $p$
is the positive square root
$p\defeq \sqrt{E^2-m^2}$. $\Omega$ is a collective notation for the
angle coordinates $(\theta,\phi)$.
$f_l$ denotes a spherical Bessel function of order $l$. We will
consider spherical Bessel functions of the first kind, denoted $j_l$,
and of the second kind, denoted $n_l$. The former describe globally
defined solutions, while the latter describe solutions that are
singular at the origin. We will also employ the
spherical Bessel functions of the third kind (or Hankel functions)
$h_l\defeq j_l+\im n_l$ and $\overline{h_l}\defeq j_l -\im n_l$.

$Y^m_l$ denotes the spherical harmonic defined
through the associated Legendre function $P^m_l$ via
\[
 Y^m_l(\theta,\phi)\defeq \sqrt{\frac{(2l+1) (l-m)!}{4\pi (l+m)!}}
  P^m_l(\cos\theta) e^{\im m \phi} .
\]
Note that complex conjugation yields $\overline{Y^m_l}=Y^{-m}_l$. The
spherical harmonics satisfy the orthogonality relation
\[
\int\xd\Omega\, Y^m_l \overline{Y^{\hat{m}}_{\hat{l}}}
 =\delta_{l,\hat{l}}\delta_{m,\hat{m}} .
\]
Here $\xd\Omega\defeq\frac{1}{4\pi}\xd\phi\xd\theta\sin\theta$.
We also remark that
\begin{equation}
 (\lap_r f_l)(pr)=\left(-p^2+\frac{l(l+1)}{r^2}\right) f_l(pr),\quad
 (\lap_\Omega Y^m_l)(\Omega)=-\frac{l(l+1)}{r^2} Y^m_l(\Omega) ,
\label{eq:spheval}
\end{equation}
where $f_l$ is any of the spherical Bessel functions.

As in the case of timelike hyperplanes the space of physical field
configurations on a hypercylinder is the space of only those
configurations that extend to a classical solution. From $|E|\ge m$ we
can infer that the eigenvalues of the operator $-\partial_0^2$ on the
physical configuration space must be larger or equal to $m^2$. Note
that we parametrize all hypercylinders in the same way, irrespective
of radius, namely via the solid angle $\Omega$ and the time $t$.

\subsection{Propagators}
\label{sec:hcprop}

We shall be interested in two types of propagation regions: the solid
hypercylinder $B_R$ and the region $B_{[R,\hatR]}$ between two
hypercylinders. In both
cases we wish to evaluate the path integral (\ref{eq:prop}) in the
same way used in the cases of hyperplanes, namely using a classical
solution matching the boundary data. This
entails a seeming contradiction. Namely, if the field configuration on
a single hypercylinder is in one-to-one correspondence to classical
solutions then the combination of field configurations on two
hypercylinders cannot be in one-to-one correspondence to classical
solutions and vice versa. This apparent contradiction has the
following resolution. We only require the classical solutions to be
defined within the propagation region. This implies that for the
solid hypercylinder $B_R$ we are restricted to the solutions
(\ref{eq:sphsol}) with spherical Bessel functions of the first kind.
In contrast, the region $B_{[R,\hatR]}$ does
not contain the time axis. Hence, in addition we may admit solutions
arising from spherical Bessel functions of the second kind.
We shall see that this somewhat heuristic procedure leads to
a consistent picture.

The propagator for the full hypercylinder is the easier one to work
out. Note that the boundary version (\ref{eq:actionbdy}) of the action
on a classical solution now takes the form
\begin{equation}
 S_R(\phicl)=-\frac{1}{2}\int\xd t\,\xd\Omega\, 4\pi R^2
 \phicl(t,R,\Omega)(\partial_r\phicl)(t,R,\Omega) .
\label{eq:bdyacthcyl}
\end{equation}
Combining this with (\ref{eq:sphsol}) using the spherical Bessel
functions of the first kind yields the propagator
\begin{equation}
 Z_R(\varphi)=N_R \exp\left(-\frac{1}{2}\int\xd t\,\xd\Omega\, 4\pi
 \varphi(t,\Omega) \im p R^2 \frac{j_l'(pR)}{j_l(pR)} \varphi(t,\Omega)
\right) .
\label{eq:prophcyl}
\end{equation}
Here $j_l'$ denotes the derivative of $j_l$. The expression
\[
p R^2 \frac{j_l'(pR)}{j_l(pR)}
\]
is to be understood as an operator defined through its eigenvalues on
a mode expansion of the field
configuration. To this end note that $p=\sqrt{E^2-m^2}$ can be
extracted from the temporal plane wave mode expansion while $l$ can be
extracted from the spherical harmonic mode expansion.

Consider now the region $B_{[R,\hatR]}$ between nested hypercylinders.
The boundary form of the action is the
difference of two terms of the form (\ref{eq:bdyacthcyl}).
To obtain a propagator we may start by splitting a classical solution
into two components. For example, in terms of a regular and a singular
component we have,
\[
\phicl(t,r,\Omega)=j_l(pr) \varphi_\text{reg.}(t,\Omega) +
 n_l(pr) \varphi_\text{sing.}(t,\Omega) . 
\]
Here $j_l(pr)$ and $n_l(pr)$ are understood as operators in the sense
described above. The radial derivative yields
\[
(\partial_r\phicl)(t,r,\Omega)= p\, j_l'(pr) \varphi_\text{reg.}(t,\Omega) +
 p\, n_l'(pr) \varphi_\text{sing.}(t,\Omega) .
\]
Using this and inverting the formal linear transformation
\[
 \begin{pmatrix}\varphi \\ \hat{\varphi}\end{pmatrix}
 =\begin{pmatrix} j_l(pR) & n_l(pR)\\
 j_l(p\hatR) & n_l(p\hatR)\end{pmatrix}
 \begin{pmatrix}\varphi_\text{reg.} \\ \varphi_\text{sing.}
 \end{pmatrix}
\]
leads to the propagator
\begin{equation}
 Z_{[R,\hatR]}(\varphi,\hat{\varphi})=N_{[R,\hatR]}
 \exp\left(-\frac{1}{2} \int\xd t\,\xd\Omega\, 4\pi
 \begin{pmatrix}\varphi & \hat{\varphi} \end{pmatrix} W_{[R,\hatR]}
 \begin{pmatrix}\varphi \\ \hat{\varphi} \end{pmatrix}
\right)
\label{eq:proptube}
\end{equation}
with
\[
 W_{[R,\hatR]}\defeq\frac{\im p}{\delta_l(pR,p\hatR)}
 \begin{pmatrix}R^2\sigma_l(p\hatR,pR) & -\frac{1}{p^2} \\
  -\frac{1}{p^2} & \hatR^2\sigma_l(pR,p\hatR) \end{pmatrix} .
\]
The functions $\delta_l$ and $\sigma_l$ are to be understood as
operators and have the following definitions:
\begin{align*}
 \delta_l(z,\hat{z}) & =j_l(z) n_l(\hat{z})- n_l(z) j_l(\hat{z})
  =\frac{\im}{2}\left(h_l(z)\overline{h_l}(\hat{z})
 -\overline{h_l}(z)h_l(\hat{z})\right) \\
 \sigma_l(z,\hat{z}) & =j_l(z) n_l'(\hat{z}) - n_l(z) j_l'(\hat{z})
  =\frac{\im}{2}\left(h_l(z)\overline{h_l}'(\hat{z})
 -\overline{h_l}(z)h_l'(\hat{z})\right) .
\end{align*}

\subsection{Vacuum}

We now turn to the question of the vacuum state on the
hypercylinder. Again we make a Gaussian ansatz of the form
(\ref{eq:vacansatz}). The precise form is now
\[
 \psi_{R,0}^I(\varphi)=C_R \exp\left(-\frac{1}{2}\int\xd t\,\xd\Omega\,
 4\pi \varphi(t,\Omega) (B_R^I \varphi)(t,\Omega)\right) .
\]
Here $B_R^I$ denotes a family of operators indexed by the radius
$R$. Recall that a state with given physical meaning changes depending
on the orientation of the carrying hypersurface with respect to the
propagation region. Furthermore, in contrast to the case of
hyperplanes, a hypercylinder with given orientation is not related to
the hypercylinder with opposite orientation by any symmetry. Thus, we
must a priori expect the vacuum wave function to be different
for the two orientations.
We have indicated this above by the superscript $I$, with
$\psi_{R,0}^I$ being the vacuum on the inner side of the
hypercylinder. Correspondingly, we
denote by $\psi_{R,0}^O$ the vacuum on the outside and by $B_R^O$ the
associated family of operators. Of course we expect the
two vacua to satisfy (\ref{eq:vacconj}), i.e.,
to be related by complex conjugation,
$\overline{B_R^O}=B_R^I$. We also suppose that we can choose $C_R$ to
be real.

An obvious condition to be satisfied by the operators $B_R^I$ (or
$B_R^O$) is that they must be related to each other, for different
radii, by propagation via (\ref{eq:proptube}). This condition is
weaker than a full invariance condition, which was essentially
sufficient to determine the vacuum in the case of hyperplanes. We
might thus expect to require additional conditions to determine the
vacuum uniquely. In any case, the propagation condition leads to the
equation
\begin{equation}
 \left(z^2 \sigma_l(\hat{z},z)-\im p \delta_l(z,\hat{z}) B_R^I\right)
 \left(\hat{z}^2 \sigma_l(z,\hat{z})+\im p \delta_l(z,\hat{z})
  B_{\hatR}^I\right) = 1 ,
\label{eq:vaccond}
\end{equation}
with $z\defeq p R$ and $\hat{z}\defeq p \hatR$.

For the general boundary formulation to be consistent over various
topologies and geometries of hypersurfaces we would like the vacuum to
be determined ``locally'' in a suitable sense. That is, the dependence
of the vacuum wave function on the field configuration on a small piece
of a hypersurface should be independent of the global topological or
geometrical nature of the hypersurface. Furthermore, we
would like the vacuum functional to ``change smoothly'' under ``smooth
changes'' of the hypersurface.
In terms of the ansatz (\ref{eq:vacansatz}) we desire these properties of
the operator $A$. In the concrete case at hand we may use this to
demand that the operator $B_R^I$ for large radii approximates the
operator $\kappa_1$ (and its appropriately rotated versions) which was
found to describe the vacuum on timelike hyperplanes.

Concretely, consider a small region near the positive $x_1$ axis at
large fixed radius $R$, i.e., in spherical coordinates $\phi\approx 0$
and $\theta\approx\pi/2$ at $r=R$. There, $\partial_2\approx
\frac{1}{R}\partial_\phi$ and
$\partial_3\approx\frac{1}{R}\partial_\theta$. Thus,
$\partial_2^2+\partial_3^2\approx
\frac{1}{R^2}(\partial_\phi^2+\partial_\theta^2)=\lap_\Omega$.
We thus
demand
\begin{equation}
B_R^I\approx R^2\sqrt{-\partial_0^2+\lap_\Omega-m^2}
\label{eq:approxvac}
\end{equation}
for large $R$ in a suitable sense. Note that in this form the condition is
rotationally invariant, independent of our initial
consideration of a specific spatial direction (namely the $x_1$-axis).
The factor $R^2$ comes from the differently scaled
integration measures on the hyperplane versus the hypercylinder.
In terms of eigenvalues on spherical harmonics and plane temporal
waves the expression (\ref{eq:approxvac}) takes the form
\[
B_R^I\approx R^2\sqrt{p^2-\frac{l(l+1)}{R^2}}
 \xrightarrow[R\to\infty]{} p R^2 .
\]
The indicated limit is understood with respect to fixed eigenvalues
$p$ and $l$.

Our strategy is thus to take the condition (\ref{eq:vaccond}), solve
for $B_R^I$, insert $B_{\hatR}^I=p\hatR^2$ and evaluate the limit
$\hatR\to\infty$. Indeed, the limit exists and the solution is
\[
B_R^I=\frac{1+\im z^2\left(j_l(z) j_l'(z)+n_l(z) n_l'(z)\right)}
 {p\left(j_l^2(z)+n_l^2(z)\right)}
 =\frac{1+\frac{\im z^2}{2}\left(h_l(z) \overline{h_l}'(z)
 +\overline{h_l}(z) h_l'(z)\right)}
 {p\, h_l(z)\overline{h_l}(z)} .
\]
Reinserting this for $B_R^I$ and $B_{\hatR}^I$ into (\ref{eq:vaccond})
confirms that we have found an actual solution of this equation. We
remark also that, as is easy to see, there is another solution to
(\ref{eq:vaccond}) given by $-\overline{B_R^I}$ which asymptotically
approximates $-\kappa_1$, thus recovering the ambiguity encountered
earlier.

Note that $B_R^I$ turns out to be a \emph{rational} function of $z$,
without singularities for positive $z$.
Moreover, the operator $z^2 \left(j_l^2(z)+n_l^2(z)\right)$ can be
expressed in terms of $p^2$ and $\lap_\Omega$ via a sum as follows,
\[
 z^2 \left(j_l^2(z)+n_l^2(z)\right)=\sum_{k=0}^{\infty}
  \prod_{j=0}^{k-1} \frac{2j+1}{2j+2}\frac{1}{p^2}
  \left(-\lap_\Omega-\frac{j(j+1)}{R^2}\right) .
\]
This can be derived using the eigenvalues (\ref{eq:spheval}) and
suitable facts about
spherical Bessel functions, see e.g.\ \cite{AbSt:handbook}. 

Turning to the outside version of the vacuum, we observe that the
relevant propagator (from $\hatR$ to $R$) is the same as
(\ref{eq:proptube}) except for the overall sign of the exponent. Since
the exponent is purely imaginary this corresponds to a complex
conjugation and the resulting relation between $B_R^O$ and
$B_{\hatR}^O$ is simply the complex conjugate of
(\ref{eq:vaccond}). Consequently, $B_R^O\defeq \overline{B_R^I}$
solves the condition. Furthermore, it obviously has the same asymptotic
limit as $B_R^I$ and is thus the required outside vacuum. Hence, the
vacuum satisfies the conjugation condition (\ref{eq:vacconj}) as
expected.
One may
speculate that the imaginary parts of $B_R^I$ and $B_R^O$ are related
to the curvature of the hypersurface.

The normalization condition (\ref{eq:vacnorm}) for the vacuum yields
\begin{equation*}
 |C_R|^{-2}= \int\xD\varphi\,
 \exp\left(-\frac{1}{2}\int\xd t\,\xd\Omega\,
  4\pi \varphi(t,\Omega) \left(\frac{2}{p\,h_l(p R)\overline{h_l}(p R)}
  \varphi\right)(t,\Omega)\right) .
\end{equation*}
In turn we can use this to determine the normalization of the
propagators. By condition (\ref{eq:vacampl}) the contraction of the
outside vacuum with the full
hypercylinder propagator (\ref{eq:prophcyl}) should give one. This
implies
\begin{equation*}
 N_R^{-1}= \overline{C_R}^{-1} \int\xD\varphi\,
 \exp\left(-\frac{1}{2}\int\xd t\,\xd\Omega\,
 4\pi \varphi(t,\Omega) \left(\frac{1}{p\, j_l(p R) h_l(p R)}
  \varphi\right)(t,\Omega)\right) .
\end{equation*}
For the normalization of the propagator for the nested hypercylinders
we obtain
\begin{multline*}
 N_{[R,\hatR]}^{-1}= C_R C_{\hatR}^{-1}\\ \int\xD\varphi\,
 \exp\left(-\frac{1}{2}\int\xd t\,\xd\Omega\,
 4\pi \varphi(t,\Omega) \left(\frac{\im\, \overline{h_l}(p \hatR)}
 {p\, \overline{h_l}(p R)\delta_l(p R,p \hatR)}
  \varphi\right)(t,\Omega)\right) .
\end{multline*}

One can now check that the unitarity condition (\ref{eq:propunit}) is
satisfied for both types of propagators. Furthermore, the composition
property (\ref{eq:pintcomp}) is satisfied both for composing
(\ref{eq:proptube}) with itself as well as for composing
(\ref{eq:proptube}) with (\ref{eq:prophcyl}). As in the case of
timelike hyperplanes, the restriction of configuration spaces makes
the validity of the composition rule a non-trivial result.

\subsection{Particle states}

We now turn to particle states. As in the case of timelike
hyperplanes, we present here a purely technical discussion, postponing
the interpretational questions to Section~\ref{sec:interpret}.

We start with the one-particle state. We consider
first an ``outside'' state. The state may be characterized in terms of
the mode expansion, i.e., through its energy $E$ and
angular momentum ``quantum numbers'' $l$ and $m$. Set
\begin{equation}
 \varphif^{\pm}_{R,l,m}(E)= \int\xd t\,\xd\Omega\, 4\pi
 \frac{\sqrt{2}}{\sqrt{p}|h_l(pR)|}
 e^{\pm \im E t} Y^{\mp m}_l(\Omega)
 \varphi(t,\Omega) .
\label{eq:sphfourier}
\end{equation}
As in (\ref{eq:tfourier}) we set $E>0$ and encode the sign of the
energy through a separate index. With this a one-particle state
reads
\[ 
 \psi^{O,\pm}_{R,E,l,m}(\varphi)=\varphif^{\pm}_{R,l,m}(E)
 \psi_{R,0}^O(\varphi) .
\]
The inner product is given by
\begin{equation}
\langle \psi^{O,a}_{R,E,l,m},\psi^{O,a'}_{R,E',l',m'}\rangle^O_R
 = 8\pi^2\delta(E-E') \delta_{l,l'}\delta_{m,m'}\delta_{a,a'}.
\label{eq:iprodhc}
\end{equation}

The corresponding ``inside'' state is
the complex conjugate, i.e.,
\begin{equation}
 \psi^{I,\pm}_{R,E,l,m}(\varphi)
 =\overline{\psi^{O,\pm}_{R,E,l,m}(\varphi)}
 =\overline{\varphif^{\pm}_{R,l,m}(E)\psi_{R,0}^O(\varphi)}
 =\varphif^{\mp}_{R,l,m}(E)
 \psi_{R,0}^I(\varphi) .
\label{eq:conjhc}
\end{equation}
The inner product is the same as (\ref{eq:iprodhc}), with index $O$
replaced by index $I$.

An ``outside'' two-particle state is given by
\begin{multline*}
 \psi^{O,a,a'}_{R,(E,l,m),(E',l',m')}(\varphi)=\\
 \left(\varphif^{a}_{R,l,m}(E)\varphif^{a'}_{R,l',m'}(E')
 -8\pi^2 \delta(E-E') \delta_{l,l'}\delta_{m,m'}\delta_{a,-a'}\right)
 \psi_{R,0}^O(\varphi) .
\end{multline*}
Multi-particle states are obtained in analogy to the procedure in the
case of hyperplanes, i.e., by starting with the suitable monomial
times the vacuum as the wave function and projecting out particle
states of lower particle number.

Amplitudes may now be associated with two different types of regions.
On the one hand, we have
``transition'' amplitudes between hypercylinders of different radii
via the propagator (\ref{eq:proptube}). This is somewhat analogous to
the propagation between parallel hyperplanes although with the
difference that the two hypersurfaces in question are not isometric.
On the other hand we have the conceptually novel
possibility of considering amplitudes for a single
hypercylinder via its solid propagator (\ref{eq:prophcyl}). This
amplitude \emph{cannot} be written as a ``transition'' amplitude
between hypersurfaces in the sense of (\ref{eq:indevol}).

The boundary of the region $B_{[R,\hatR]}$ consists of two
hypercylinders. As boundaries they are oriented. Concretely, the
smaller hypercylinder of radius $R$ is oriented ``inside'' and the
larger hypercylinder of radius $\hatR$ is oriented ``outside''. Given
an ``inside'' one-particle state at radius $R$ and an ``outside''
one-particle state at radius $\hatR$ yields
\begin{equation}
 \rho_{[R,\hatR]}(\psi^{I,a}_{R,E,l,m},\psi^{O,a'}_{\hatR,E',l',m'})
 =\alpha_{[R,\hatR],p,l} 8\pi^2\delta(E-E')
  \delta_{l,l'}\delta_{m,m'}\delta_{a,a'}.
\label{eq:hnampl}
\end{equation}
Here,
\[
 \alpha_{[R,\hatR],p,l}\defeq
 \frac{h_l(p\hatR)}{h_l(pR)}\frac{|h_l(pR)|}{|h_l(p\hatR)|}
 =\frac{\overline{h_l}(p R)}{\overline{h_l}(p \hatR)}
 \frac{|h_l(p\hatR)|}{|h_l(p R)|} .
\]

As an example of an amplitude for the solid hypercylinder using
(\ref{eq:prophcyl}) we evaluate the two-particle state shown above,
\begin{equation}
 \rho_R(\psi^{O,a,a'}_{R,(E,l,m),(E',l',m')})=
 \frac{h_l(p R)}{\overline{h_l}(p R)}8\pi^2 \delta(E-E')
 \delta_{l,l'}\delta_{m,m'}\delta_{a,-a'}.
\label{eq:hcampl}
\end{equation}
Note that this type of amplitude is defined exclusively for
``outside'' states.

\section{Interpretation}
\label{sec:interpret}

So far we have dealt with the formal side of the general boundary
formulation showing that a consistent picture of states, vacua,
amplitudes etc.\ emerges for the hypersurfaces and regions
considered. In terms of the companion paper \cite{Oe:GBQFT}, all the
core axioms as well as the vacuum axioms are satisfied.
We now turn to the physical interpretation of those
structures.

\subsection{Particles on timelike hypersurfaces}

An initial discussion of particle states on timelike hypersurfaces was
already given in \cite{Oe:timelike} (in the hyperplane case). We
review parts of this discussion here and add the novel aspects arising
from the hypercylinder case.

A crucial difference between states on spacelike and timelike
hypersurfaces arises as follows. In the spacelike case
causality implies that a state is purely an incoming state or
an outgoing state depending on whether it forms the beginning or the
end of a time-evolution process. This distinction is encoded in the
standard formulation by whether the state is a ket-state (in-state) or a
bra-state (out-state). We have linked this distinction in
Section~\ref{sec:sl} to the orientation (past or future) of the
carrying hypersurface. 
In contrast, in the timelike case, a state on a
given oriented hypersurface is neither necessarily an incoming state
nor an outgoing one. Rather, each particle within
the state may be independently incoming or outgoing. This choice is
precisely given by the sign of the energy in (\ref{eq:tfourier}) and
(\ref{eq:sphfourier}), encoded by the index $\pm$. We set the negative
sign to represent in-particles and the positive sign to represent
out-particles. (This makes momenta on spacelike and timelike
hyperplanes mutually consistent \cite{Oe:timelike}.)

In the case of particles on hypercylinders the parametrization of
particle states we have chosen makes immediate sense, thinking in
terms of classical waves expanded spherically. In particular, (away from the
center) it is natural to think in terms of incoming or outgoing
waves. In the hyperplane case we are more accustomed to think in terms
of a plane wave expansion parametrized by 3-momenta $p$. However, the
parametrization in terms of energy $E$, incoming versus outgoing $\pm$ and
2-momentum $\tilde{p}$ is easily related. Indeed, the equation
$E^2=p_1^2+\tilde{p}^2+m^2$ determines the missing momentum component
$p_1$ already up to a sign. This sign is now determined indirectly by
$\pm$. Namely, the momentum of an in-particle has to point into the
propagation region while that of an out-particle has to point out of
the propagation region. Hence, for an in-particle on an oriented
hyperplane $p_1$ is directed to the opposite side of the hyperplane,
while for an out-particle it is directed to the same side.

Note also that the relation to orientation change via complex
conjugation (\ref{eq:conj}) is consistent. Namely, an in-particle
considered on a hyperplane with opposite orientation must become an
out-particle since the propagation region is now on the other
side. Correspondingly for out-particles. As we have seen in
(\ref{eq:conjtl}) and (\ref{eq:conjhc}), this is
precisely what happens. 

The amplitudes we have calculated are also consistent with the in/out
interpretation of particle states we have given. Recall that in the
non-interacting theory we are considering, amplitudes simply express
all the possibilities in which particles on the boundary could be
identical. In the hyperplane case, this means that an in-particle on
one hyperplane can only pair with an out-particle on the other one and
vice versa. This is exactly what we see in (\ref{eq:tampl}). The
situation is similar for two nested hypercylinders, namely an
in-particle on one hypercylinder can only pair with an out-particle on
the other one and vice versa, see (\ref{eq:hnampl}).
A different situation arises for the solid
hypercylinder. Classically, an incoming spherical wave produces an
outgoing one and vice versa. Indeed, we see in this case that
in-particles only pair with out-particles and vice versa, see
(\ref{eq:hcampl}).

\subsection{Probability interpretation}

In the standard formulation the modulus square of a transition
amplitude from a state $\psi$ to a state $\eta$ is the probability of
observing the final state $\eta$ (rather than its orthogonal
complement) given that the state $\psi$ was prepared initially.
Obviously, this interpretation is not applicable to generic situations
arising in the proposed formulation, where an amplitude may be
evaluated on a \emph{single} state space.

In cases where we have a region bounded by two disjoint
hyperplanes we can still use the standard interpretation with minimal
change. For example, consider the case of two parallel timelike
hyperplanes. The modulus square of an amplitude of the type
(\ref{eq:tampl}) provides the probability of ``observing'' a state on
one hyperplane given that another one was ``prepared'' on the other
hyperplane. (In the case at hand one state is $\psi_{E,\tilde{p}}^a$
and the other one is $\psi_{E',\tilde{p}'}^{a'}$.) Nevertheless, a crucial
difference to the standard formulation is that the definite temporal
character of the procedure is lost. That is, ``preparation'' no longer
necessarily precedes ``observation''. Rather, we are dealing with a
more general conditional probability. The amplitudes for regions
between nested hypercylinders might be interpreted similarly, namely,
as \emph{transition} amplitudes from one hypercylinder to the other
one. Correspondingly the associated probability may be interpreted
as that of one state being measured on one hypercylinder conditional
on another state being present on the other hypercylinder.

A probability interpretation for the general case is proposed in the
companion paper
\cite{Oe:GBQFT}. We briefly recall it in its general form. Let $M$ be
a region with boundary $\sig$, the associated state space being
$\cH_\sig$. We denote the amplitude by $\rho_M:\cH_\sig\to\C$.
We specify part of a measurement process through a closed
subspace $\cS\subset\cH_\sig$. This may be thought of as representing
certain knowledge about the process (compare to
``preparation''). Furthermore, we specify a second closed subspace
$\cA\subseteq\cS$. This may be thought of as representing a question
posed in the process, namely, whether the state corresponding to the
measurement is in the subspace $\cA$ (rather than in its orthogonal
complement) given that it is in $\cS$ (compare to ``observation'').
Let $\{\xi_i\}_{i\in I}$ be an
orthonormal basis of $\cS$, which reduces to an orthonormal basis of
$\cA$ given by $\{\xi_i\}_{i\in J\subseteq I}$.
The probability $P(\cA|\cS)$ associated with the process is given by
the quotient,
\begin{equation}
 P(\cA|\cS)=\frac{\sum_{i\in J}|\rho_M(\xi_i)|^2}
 {\sum_{i\in I}|\rho_M(\xi_i)|^2} .
\label{eq:prob}
\end{equation}

It is shown in \cite{Oe:GBQFT} that this interpretation can be
reduced to the standard one in the standard circumstances
(time-evolution between spacelike hyperplanes). Furthermore, it covers
some less standard conditional probabilities that can be inferred from
the standard ones. Here, however, we shall be interested in a genuinely
non-standard application where the boundary of the region associated
with the measurement is connected.

Consider a hypercylinder of radius $R$. Call the associated outside
state space $\cH$. Given a function $f_{l,m}(E)$ satisfying
\begin{equation}
 8\pi^2\int\xd E\sum_{l,m} |f_{l,m}(E)|^2=1 ,
\label{eq:fnorm}
\end{equation}
the one-particle states defined by the wave functions
\[
 \psi^{O,\pm}_f(\varphi)\defeq \int\xd E\sum_{l,m} f_{l,m}(E)
 \psi^{O,\pm}_{R,E,l,m}(\varphi)
\]
are normalized due to (\ref{eq:iprodhc}). Similarly, taking another
function $f_{l,m}'(E)$ satisfying (\ref{eq:fnorm}), the two-particle
states given by
\[
 \psi^{O,a,a'}_{f,f'}(\varphi)\defeq
 \int\xd E\,\xd E'\sum_{l,m,l',m'} f_{l,m}(E) f_{l',m'}'(E')
 \psi^{O,a,a'}_{R,(E,l,m),(E',l',m')}(\varphi)
\]
are normalized.

Now remember that $\cH$ is a Fock space and may be decomposed
into a direct sum of components $\cH_n$ with given particle
number $n$, i.e., $\cH=\bigoplus_{n=0}^\infty \cH_n$. Now define a
closed subspace $\cS_f\subset\cH$ as follows,
\[
 \cS_f\defeq
 \{ \eta\in\cH_2 | \exists \lambda\in\C,f',a':
 \eta=\lambda\psi^{O,-,a'}_{f,f'}\} .
\]
Here $f'$ is supposed to be a function satisfying (\ref{eq:fnorm}). In
words this means the following: $\cS_f$ is the subspace of the space
of two-particle states where one particle is an in-particle that can
be described by the function $f$ in the sense given above. For
example,
$f$ might be peaked around a particular energy and particular angular
momentum quantum numbers, thus describing a particle wave packet with
approximately these properties.

Let the subset $\cA_{f,g}\subset\cS_f$ be spanned by the single state
$\psi^{O,-,+}_{f,g}$ for a function $g$ satisfying
(\ref{eq:fnorm}). As we will explain later, it turns out that the
denominator of (\ref{eq:prob}) is equal to one in the present case.
This implies,
\begin{equation}
P(\cA_{f,g}|\cS_f)=\left|\rho_R(\psi^{O,-,+}_{f,g})\right|^2=
\left|8\pi^2\int\xd E\sum_{l,m}
\frac{h_l(pR)}{\overline{h_l}(p R)} f_{l,m}(E) g_{l,m}(E)\right|^2
\label{eq:probfg}
\end{equation}
Note that due to the normalization (\ref{eq:fnorm}) and the fact that
$h_l(pR)/{\overline{h_l}(p R)}$ has modulus one this quantity is
less or equal than one as required.

Physically, $P(\cA_{f,g}|\cS_f)$
is the probability that we ``observe'' an outgoing particle
characterized by a wave packet determined by $g$, given that an
incoming particle with wave packet determined by $f$ was
``prepared''. Note that we could also reverse the role of the incoming
and the outgoing particle. However, the meaning of ``preparing'' and
``observing'' would be less intuitive then. It is clear that we can
extend this example to multi-particle states. That is, we can define a
subspace $\cS$ such that the total number of particles is fixed and
some of them have determined wave packets. We can then ``test'' via
$\cA$ for specific wave packets for the remaining particles.

Considering certain outgoing particles conditional on certain incoming
ones is of course what one usually does in perturbative quantum field
theory. The difference is that in the standard formulation these
particles live in different state spaces. However, we can artificially
produce a similar situation here. It turns out that the Fock space
$\cH$ may be decomposed into a product of a Fock space of incoming
particles $\cH^-$ and one of outgoing particles $\cH^+$, i.e.,
$\cH=\cH^-\tens\cH^+$. Indeed, we may construct $\cH^+$ and $\cH^-$ in
terms of the respective subspaces of $\cH$. Note that these inherit
inner products in this way and are isomorphic. Also, since incoming
and outgoing particles are mutually orthogonal, the inner product of
$\cH$ is identical to that reconstructed from those of $\cH^-$ and
$\cH^+$.

What is more, defining $\bar{\cH}^+$ to be the dual Hilbert space of
$\cH^+$ it turns out that the map $\tilde{\rho}_R:\cH^-\to\bar{\cH}^+$
induced by the amplitude is not only well defined, but preserves the
inner product. Thus, the decomposition $\cH=\cH^-\tens\cH^+$ behaves
exactly as if it was induced by a decomposition of the carrying
hypersurface. (This is also the reason why the denominator of the
probability expression (\ref{eq:probfg}) is equal to one.)
We recover a description which shows resemblance to the standard
formulation. However, the map $\tilde{\rho}_R$ is of course not simply
a ``time-evolution''.

\section{Conclusions and Outlook}
\label{sec:concl}

We hope to have presented in this work a compelling example of a
general boundary quantum field theory in the shape of the Klein-Gordon
theory. The regions and hypersurfaces considered are still much less
general than what one would like to allow (see also the discussion in
\cite{Oe:GBQFT} on this topic). However, the hypercylinder case in
particular exhibits many of the novel and non-standard features of the
general boundary formulation. This also included the first concrete
application of the generalized probability interpretation proposed in
\cite{Oe:GBQFT} in a context beyond the reach of standard quantum
mechanics. 

Note that the Fock space structure of the state space was instrumental
in the probability interpretation of the solid
hypercylinder example. It allowed to construct subsets of the state
space with a clear physical meaning is a simple way. It might be
expected that this will be a much more difficult problem in genuinely
non-perturbative theories, where no convenient grading of the state
space is available. For example, in the case of gravity, it is a
priori highly unclear which properties of the (quantum) geometry of a
hypersurface we may set fixed and for which sub-properties we may
then meaningfully ``ask'' in a measurement process.

The Klein-Gordon theory is obviously only a starting point and more
complicated quantum field theories should be considered. However, we
expect that for free theories this should be relatively
straightforward. See, e.g., the discussion of spinor and gauge fields
in the Schr\"odinger representation in \cite{Jac:schroedinger}. For
interacting theories, we expect that the usual perturbative approach
can be carried over. In particular, it should be possible to derive
the S-matrix through an infinite radius limit of the solid
hypercylinder amplitude. Indeed, one might argue that this would be
conceptually more
satisfactory than the usual derivation from equal-time hyperplanes at
large negative and positive times. Namely, considering the
interaction to be negligible at large distances in space (from the
experiment) appears more natural than considering it negligible at
very early or late times. In particular, this would be compatible with
genuinely static processes. Note that such a derivation would make the
crossing symmetry of the S-matrix \emph{manifest}, since incoming and
outgoing particles are part of a single state. Indeed, precisely for
this reason, crossing symmetry was taken in
\cite{Oe:catandclock} as a strong indication for the validity of
the general boundary formulation.

\bibliography{stdrefs}
\bibliographystyle{amsordx}

\end{document}